\documentclass[pra,twocolumn,showpacs,floatfix]{revtex4}

\usepackage{amssymb,amsmath,amsfonts}
\usepackage{color}
\usepackage{graphicx}

\newcommand{\average}[1]{\langle#1\rangle}
\newcommand{\pderiv}[2]{\frac{\partial#1}{\partial#2}}
\newcommand{\pderivtwo}[2]{\frac{\partial^2#1}{\partial#2^2}}

\begin{document}

\bibliographystyle{prsty}

\title{Excitation spectrum of bosons in a finite one-dimensional circular waveguide via the Bethe ansatz}

\author{Andrew G. Sykes}
\email{sykes@physics.uq.edu.au}
\author{Peter D. Drummond}
\author{Matthew J. Davis}

\affiliation{ARC Centre of Excellence for Quantum-Atom Optics, School of Physical Sciences, University of Queensland, Brisbane, QLD 4072, Australia}

\date{\today}

\begin{abstract}

The exactly solvable Lieb-Liniger model of interacting bosons in  one-dimension
has attracted renewed interest as current experiments with  ultra-cold atoms
begin to probe this regime.  Here we numerically solve  the equations arising
from the Bethe ansatz solution for the exact  many-body wave function in a
finite-size system of up to twenty particles for attractive interactions.  We
discuss the novel features  of the solutions, and how they deviate from the
well-known string solutions  [H. B. Thacker, Rev. Mod. Phys.\ \textbf{53}, 253
(1981)] at finite densities. We present excited state string solutions in the
limit of  strong interactions and discuss their physical interpretation, as well
as the characteristics of the quantum phase transition that occurs as a function
of interaction strength in the mean-field limit.    Finally  we compare our results to those of exact
diagonalization of the many-body  Hamiltonian in a truncated basis.  
We also present excited state solutions and the excitation
spectrum  for the repulsive 1D Bose gas on a ring.

\end{abstract}

\maketitle

\section{Introduction}

Physics in low dimensional systems has long provided a rich source of fascinating and often unexpected phenomena. With the steady progress of experimental methods in ultra-cold gases, {effective one-dimensional} systems are beginning to be realised in the laboratory~\cite{bec_low_dim,bill,tonksgas,1dbox,dalibard1,foot1,bec_one_dim,greiner_low_dim_bec,bec_low_dim2,bec_low_dim3}. The integrability of certain many-body problems in one dimension~\cite{liebliniger,mcguire} provides an opportunity to reliably examine many-body quantum physics beyond mean-field theory. It is becoming clear that many condensed matter theories of phase transitions and collective excitations depend on the number of degrees of freedom within the system~\cite{sachdev,greiner_qpt}. 

In this paper we focus on a system of identical bosons tightly confined in a ring trap such that the system can be considered to be purely one-dimensional (but with three-dimensional scattering). We are primarily interested in obtaining the excitation spectrum of the system. The motivation for this work was provided by the prediction of a quantum phase transition in the regime of attractive interactions. As the interaction strength for a fixed number of particles becomes more negative, zero temperature quantum fluctuations eventually cause the gas to form a soliton-like localised state~\cite{kavoulakis,kanamoto2003a}. Quantum solitons in 1D Bose gases were first predicted and observed with photons in optical fibres~\cite{drummond_soliton1,drummond_soliton2}, and more recently soliton-like behaviour has been observed in systems of massive particles with attractive interactions in quasi-1D \cite{soliton,soliton_train} and 3D geometries~\cite{soliton2}. 

The quantum phase transition in the 1D toroidal geometry was initially identified using a mean-field approach by Kavoulakis~\cite{kavoulakis} and Kanamoto \emph{et. al.}~\cite{kanamoto2003a,kanamoto2003b}. Further work by Kanamoto \emph{et al.} has taken a quantum many-body approach using exact diagonalization by truncating the Hilbert space to three (or five) single-particle states for up to 200 atoms~\cite{kanamoto1,kanamoto2}.  They found a number of interesting physical features in the region of the phase transition, in particular evidence of symmetry-breaking as the difference in energy between the ground and first excited stated tended to zero and scaled as $N^{-1/2}$.

Whilst the results obtained by Kanamoto \emph{et al.}~\cite{kanamoto1,kanamoto2} remain qualitatively correct at low temperature and weak interactions, the Bethe ansatz provides us with an exact solution for all excited states of the system at all values of interaction strength. 
The ground state quasi-momenta and energy {for the one-dimensional Bose gas has previously been} calculated using { the Bethe ansatz by Sakmann \emph{et al.}}~\cite{sakmann}.  However, we emphasize that for {experimentally realistic temperatures for this system} excited states will be involved in both the static and dynamic properties of the system. {In this paper we extend the work of Sakman \emph{et al.}~\cite{sakmann} to calculate the low-lying excitation spectrum of the finite one-dimensional Bose gas as a function of interaction strength for up to $N=20$ particles.
This is a non-trivial result due to the complicated behaviour of the quasi-momenta in the complex plane in the attractive case (see section \ref{quasi}).} 
To the best of our knowledge this work is the first example of calculations made on excited states of this system (with more than 3 particles) without any truncation of the Hilbert space and without restricting oneself to the limiting \emph{string} solutions.

The string solutions arise in the case where the interatomic interactions are sufficiently attractive, or alternatively in the zero density limit. The problem reduces to one solved by McGuire~\cite{mcguire} and elaborated upon in references~\cite{salasnich1,calogero_degasperis}. The point of interest in our numerical solutions is the deviations of the quasi-momenta from these previously known solutions. Our present work extends our understanding of quantum solitons  from the large boson numbers in optical fibres towards the much smaller numbers  possible in atomic systems.

Related work has been performed by Oelkers and Links~\cite{oelkers_bosehubbard}, who concentrate on a toroidal lattice governed by the Bose-Hubbard Hamiltonian with periodic boundary conditions. Physically one expects the results of their calculations to tend toward those of the continuum as the number of lattice sites increases. 
The Bose-Hubbard Hamiltonian is, however, non-integrable and it has been shown by Seel \emph{et al.}~\cite{seel_xxz} that the spin-1/2 XXZ Heisenberg chain (which is integrable) maps onto the 1D Bose gas model in the continuum limit. Other examples of lattice Hamiltonians that have the 1D Bose gas in the continuum limit can be found in ~\cite{lattice_1D_bosegas1,lattice_1D_bosegas2}.

The structure of this paper is as follows. In section \ref{model} we review the work of Lieb and Liniger~\cite{liebliniger} in order to correctly pose the problem. In section \ref{quasisect} we discuss our method of solution and present the results, whilst also discussing its connection to earlier work by McGuire~\cite{mcguire}. In section \ref{exc_spect} we present the excitation spectrum of the system obtained from our calculations, and in section \ref{comparison} we show a comparison of our work to that of the truncated Hilbert space approach of reference~\cite{kanamoto1,kanamoto2}.  Finally we calculate the excitation spectrum for the repulsive case in section~\ref{sec:repulsive}, before concluding in section~\ref{sec:conclusions}.

\section{The Model}\label{model}

We are considering a system of $N$ bosons, each with a mass $m$, on a ring of radius $R$, in the limiting case of the radial confinement \emph{freezing} out all the transverse degrees of freedom of the atoms. The interactions are assumed to be short range, contact interactions, modelled using the delta function. With these assumptions the system will be described by the Hamiltonian (in first quantised formalism)
\begin{eqnarray}
H=\sum_{j=1}^{N}-\pderivtwo{}{\theta_j}+2c\sum_{i<j}\delta(\theta_i-\theta_j)\label{firstham}
\end{eqnarray}
where length is measured in units of $R$ and energy is measured in units of $\hbar^2/2mR^2$ ($\theta_i$ is the angular coordinate of the $i$th particle). The parameter $c$ is related to the s-wave scattering length by Eq.~\eqref{one_dim_scattering}, and quantifies the strength of the two body interactions, $c<0$ describes an attractive gas while $c>0$ describes a repulsive gas.

Since our spatial universe is of course three dimensional it is worthwhile to note the physical limits of applicability of this model. Assuming harmonic confinement, the length scale of the dimensions tranverse to the ring are  $x_0=\sqrt{\hbar/(m\omega_{x})}$ and $y_0=\sqrt{\hbar/(m\omega_y)}$.  Another important length scale is the s-wave scattering length, $a$, that represents the strength of two body interactions.  The healing length of the gas $\xi=1/\sqrt{8\pi na}$, where $n$ is the three dimensional density, reveals how sensitive the gas is to irregular features in the potential. For the system in which we are interested, it is possible to define a \emph{local} set of cartesian coordinates such that $z_0\gg\xi\gg x_0,y_0$, where the $z$ axis runs tangential to ring. Essentially this amounts to a torus of radius $R$ and cross-sectional area $\pi r^2$ satisfying the condition $R\gg r$. In this way the energy level spacings for the tranverse ($x$ and $y$) degrees of freedom supersede all other energy scales within the system and the system is essentially one dimensional.  Therefore we can assume a ground harmonic oscillator state in the transverse dimensions, and integrate them out to 
obtain an  effective one dimensional interaction~\cite{olshanii_1d_scattering}
\begin{eqnarray}
c=\frac{\hbar^2}{m}\frac{a}{x_0y_0},\label{one_dim_scattering}
\end{eqnarray}
where $m$ is the mass of a single particle. Some interesting work has recently been done by Parola et. al.~\cite{salasnich,salasnich3} on the quasi-one dimensional limit which indicated a transverse collapse (in addition to the angular collapse) at a certain critical interaction strength. In this work they took into closer consideration the effects of the excluded two dimensions.
Current experimental realisations of such ring traps~\cite{ring_trap_theory} are still far from the quasi-1D regime~\cite{bec_in_ring,bec_in_ring2,ring_trap,ion_ring_trap,lattice_ring_trap1}. 

Given the Hamiltonian \eqref{firstham}, Schr\"{o}dinger's equation is
\begin{eqnarray}
H\psi_n(\theta_1,\ldots,\theta_N)=E_n\psi_n(\theta_1,\ldots,\theta_N),
\label{schrodinger}
\end{eqnarray}
where $n$ is the label for different eigenstates. Equation~\eqref{schrodinger} can be solved exactly via the Bethe ansatz~\cite{liebliniger}. Due to the symmetry under permutation of particle coordinates, the region over which we need to integrate Eq.~\eqref{schrodinger} can be restricted to $0\leq\theta_1\leq\ldots\leq\theta_N\leq2\pi$. Lieb and Liniger noted that in this region the delta function vanishes everywhere except along the boundaries where $\theta_i=\theta_{i+1}$, hence the problem could be cast in a different manner by writing the interactions as a \emph{boundary condition} on the restricted region rather than an explicit term in the Hamiltonian~\cite{liebliniger}. The new problem was to find the appropriate solution to 
\begin{eqnarray}
&-\sum_{j=1}^N\pderivtwo{\psi_n}{\theta_j}=E_n\psi_n\quad\textrm{in the restricted region},&\label{helmholtz}\\
&\left(\pderiv{\quad}{\theta_{j+1}}-\pderiv{\quad}{\theta_j}\right)\psi_n\vert_{\theta_{j+1}=\theta_j}=c\psi_n\vert_{\theta_{j+1}=\theta_j}&\label{interaction}
\end{eqnarray}
where Eq.~\eqref{interaction} is the  previously mentioned boundary condition arising from the interaction. Furthermore the boundary conditions pertaining to the ring geometry of the system, i.e. $\psi_n(\ldots,\theta_i,\ldots)=\psi_n(\ldots,\theta_i+2\pi,\ldots)$, must also be recast onto the new region, vis
\begin{subequations}
\label{bc}
\begin{eqnarray}
\psi_n(0,\theta_2,\ldots,\theta_N)&=&\psi_n(\theta_2,\ldots,\theta_N,2\pi),
\label{bc1}\\
\pderiv{}{\theta}\psi_n(\theta,\theta_2,\ldots,\theta_N)\vert_{_{\theta=0}}&=&\pderiv{}{\theta}\psi_n(\theta_2,\ldots,\theta_N,\theta)\vert_{_{\theta=2\pi}}.
\label{bc2}
\end{eqnarray}
\end{subequations}
The Bethe ansatz is employed as a means of integrating Eq.~\eqref{helmholtz}. The (unnormalized) wave function in the restricted region is given the form
\begin{eqnarray}
\psi_n(\theta_1,\ldots,\theta_N)=\sum_{\left\{Q\right\}}A^{(n)}_Qe^{i\sum k^{(n)}_{Q(j)}\theta_j},
\label{ansatz}
\end{eqnarray}
where $Q$ is some permutation of the integers $1,2,\ldots,N$, $Q(j)$ is the $j$th element in $Q$ and the sum runs over all the $N!$ distinct permutations of $Q$. Equation \eqref{ansatz} is one form of the Bethe ansatz and was originally used as a means of integrating spin chain Hamiltonians~\cite{bethe}. The coefficients of each permutation, $A^{(n)}_Q$, can be determined (up to some arbitrary phase) by the interaction boundary condition Eq.~\eqref{interaction}. We choose the convention $A^{(n)}_{12\ldots N}\equiv1$ and for any other permutation $Q$, $A^{(n)}_Q$ is found by constructing $Q$ out of $12\ldots N$ by transposing adjacent elements. Then $A^{(n)}_Q$ is a product of the terms
\begin{eqnarray}
-\frac{k^{(n)}_{Q(a)}-k^{(n)}_{Q(b)}+ic}{k^{(n)}_{Q(a)}-k^{(n)}_{Q(b)}-ic},\nonumber
\end{eqnarray}
where $Q(a)$ and $Q(b)$ are the adjacent elements being transposed, with $Q(a)$ to the left of $Q(b)$ \emph{before} the transposition. 

The scalar quantities $k^{(n)}_1,\ldots,k^{(n)}_N$, known as the \emph{quasi-momenta} (and sometimes the rapidities), are determined by the ring geometry of the system, i.e., by the boundary conditions in Eqs.~\eqref{bc}. These quasi-momenta furnish important quantities for the system such as the energy eigenvalues and the total momentum, however it should be stressed that (other than the case of no interactions $c=0$) the individual quasi-momenta do not have a direct relationship to the single particle momentum states. The latter must be obtained from the single particle reduced density operator, which can be obtained from the many body wave function.

Substitution of Eq.~\eqref{ansatz} into the boundary conditions \eqref{bc1} and \eqref{bc2}, yield the following set of $N$ simultaneous equations to determine the quasi-momenta as functions of $c$
\begin{eqnarray}
(-1)^{N-1}e^{-i2\pi k_j}=\prod_{s=1}^N\frac{k_j-k_s+ic}{k_j-k_s-ic},\quad j=1,\ldots,N.\label{betheeqns}
\end{eqnarray}
The purpose of this work is to explore solutions of these equations to reveal key features of the system such as the excitation spectrum of the gas as a function of the interaction strength. Analytical solutions of Eq.~\eqref{betheeqns} have previously been found for up to three particles~\cite{mugasnider}, however to our knowledge the only numerical calculations have been for the ground state energy~\cite{sakmann}. Once Eqs.~\eqref{betheeqns} have been solved it is true that, in principle at least, one has obtained all neccessary information to obtain the wave function for the system. However in practice, even for a modest number of particles such as ten, the wave function itself will involve a summation over $10!\approx4\times10^6$ terms, a somewhat cumbersome object. However, for certain physical quantities, simple analytical expressions involving only the quasi-momenta can  easily be derived, for instance the energy of the $n$th eigenstate is given by
\begin{eqnarray}
E_n=\sum_{j=1}^N{k^{(n)}_j}^2,\label{energy}
\end{eqnarray}
and likewise the total momentum by
\begin{eqnarray}
P_n=\sum_{j=1}^N{k^{(n)}_j}.\label{momentum}
\end{eqnarray}
We will utilise equation \eqref{energy} in section \ref{exc_spect} to obtain the excitation spectrum.
\begin{figure*}
\includegraphics[width=17cm]{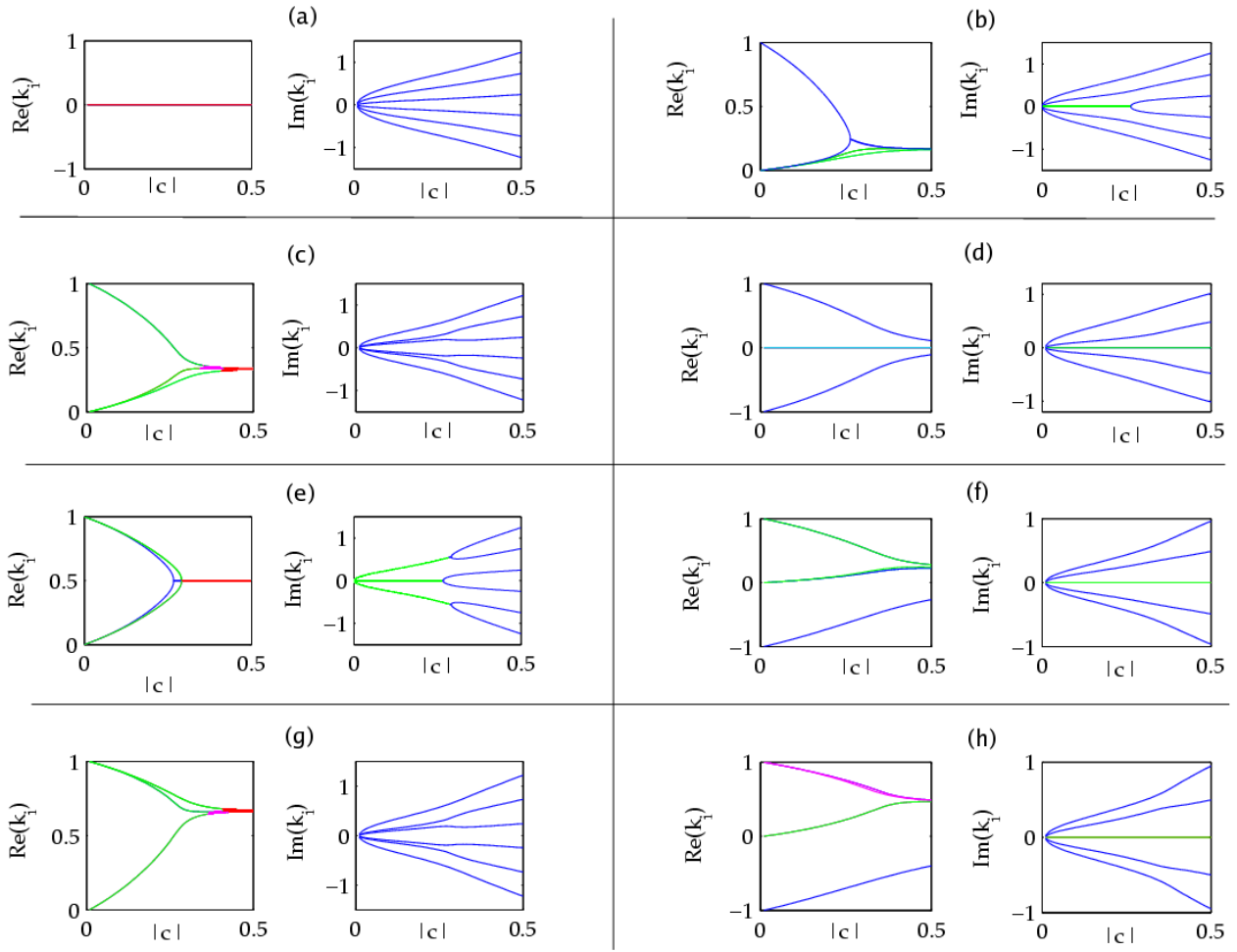}
\caption{(colour online) The behaviour of the quasi-momenta for $N=6$ particles as a function of the interaction strength $c<0$, for the first 8 states, ordered via their energy at $c=0$. On each graph there are six lines, the color of which indicates how many quasi-momenta are overlapping. Blue indicates one root, green indicates two roots, magenta indicates three roots, cyan indicates 4 roots and red indicates six roots. 
The single particle momentum states at $c=0$ are, (a) $\left\{0,0,0,0,0,0\right\}$, (b) $\left\{1,0,0,0,0,0\right\}$, (c) $\left\{1,1,0,0,0,0\right\}$, (d) $\left\{1,-1,0,0,0,0\right\}$, (e) $\left\{1,1,1,0,0,0\right\}$, (f) $\left\{1,1,-1,0,0,0\right\}$, (g) $\left\{1,1,1,1,0,0\right\}$ and (h) $\left\{1,1,1,-1,0,0\right\}$. The states with a non-zero \emph{total} angular momentum (that is states (b),(c),(e),(f),(g) and (h)) have all been chosen to have \emph{positive} total angular momentum and naturally they have a degenerate counterpart with a negative total angular momentum which can be found easily via the mapping $k_i\rightarrow-k_i$ for all $i$.}
\label{sixpartkatt}
\end{figure*}

\section{The Attractive Gas: $c<0$}\label{attractive}

The behaviour of the Bose gas with attractive interactions has received far less scrutiny than its repulsive counterpart. Originally Lieb and Liniger did not {consider} this regime to be of physical relevance~\cite{liebliniger}, however McGuire showed~\cite{mcguire} that once an $N$-particle bound state had formed, the ground state energy of the system scales as
\begin{eqnarray}
E_0 \propto -c^2N(N^2-1).
\end{eqnarray}
This posed problems for the solution in the thermodynamic limit where $N\rightarrow\infty$, $R\rightarrow\infty$ and $N/R=$ constant. In contrast for the repulsive case it is found $E_0\propto N\propto R$, and hence an expression for the energy density could be found~\cite{yang1,yangyang1,korepinbook,takahashi}. However $E/R$ diverges in the attractive regime. It is perhaps worth noting that the limit of an extremely dilute gas is free from these divergences, provided one took the limit as $N^3/R=$ constant. Another possibility is to consider the limit $R\rightarrow\infty$ while $N$ remains constant~\cite{mcguire,calogero_degasperis,calabrese_caux1,calabrese_caux2}.

As previously mentioned the quantum phase transition to a soliton-like state provides added interest for the attractive gas. This transition spontaneously breaks the translational symmetry of the system even at zero temperature. We discuss this point further in section \ref{quasi} and see how the eigenstates obtained from the Bethe ansatz maintain the {translational} symmetry of the Hamiltonian by forming a continuous superposition of localised states around the ring. We have included in appendix \ref{meanfieldc0} a simple derivation (following the work of Kavoulakis~\cite{kavoulakis}) of the critical interaction strength ($C_0=-\pi/2N$) at which the phase transition is predicted to occur.

\subsection{The Quasi-Momenta}\label{quasi}\label{quasisect}

To find the quasi-momenta  $\{k_i\}$ that fully characterise the exact solution, we must numerically solve Eqs.~\eqref{betheeqns}.  For $N$ particles this gives $N$ simultaneous nonlinear equations.  The nature of most root-finding algorithms is such that it requires a reasonable initial guess for the location of the roots.  We have {developed} a suitable procedure to solve these equations for a range of interaction strengths $c$ starting from the known solutions for the ground and excited states at $c=0$, which are given by all $k_i$ being an integer.

To find the solution for non-zero $c$, we first choose a value for $c$ close to zero, and use the ideal gas solution for the state we are interested in as the initial guess for the root-finding algorithm.  We make use of the built-in root-finder \verb|fsolve| in the software package MATLAB on a standard desktop PC.  By choosing a small enough $c$ the initial guess for the quasi-momenta are close to the actual solution, and the root-finding algorithm converges relatively rapidly
\footnote{It is necessary to use small random numbers to keep the quasi-momenta distinct from one another.}.
 
Once the first solution close to $c=0$ is found, $c$ is decreased in small increments $\Delta c$ and  an initial guess for the quasi-momenta for the new value of $c$ is based on a smooth extrapolation from the previous values. Typically the extrapolation is based only on the two previous values of $c$ and is thus linear.  Difficulties arise in this method when a particular quasi-momenta at an interaction strength $c$ differs greatly from that at $c+\Delta c$. This problem arises particularly when $c$ is close to the critical interaction strength $c_0$ and we discuss how we deal with the problem later.

The actual size of $\Delta c$ will be dependent on how large $N$ is. The scaling behaviour of the critical interaction strength goes as $C_0\propto1/N$ and we find that the efficiency of the algorithm is improved by using $\Delta c$ as small as $10^{-3}/N$. It is possible to use a larger value of $\Delta c$ but we have found that smaller step sizes give better initial predictions for the roots for the next step and results in faster convergence of the root-finding algorithm.

We have not found it necessary to make use of high precision arithmetic used by Sakmann \emph{et al.} \cite{sakmann} in finding our solutions. In their work they make use of a particular transformation [see Eq. (33) in Ref.~\cite{sakmann}] popularised in the seminal work of Lieb and Liniger~\cite{liebliniger}. The transformation simplifies the numerics in the  repulsive case by \emph{spreading} the transformed quasi-momenta  into numbers which can be easily distinguished by machine precision. The same is not true in the case of attractive interactions. As pointed out by Sakmann \emph{et al.}, working with the transformed quasi-momenta in the attractive case requires numerical precision of approximately $10^{-85}$. In this work we solve Eqs.~\eqref{betheeqns} directly. We set the tolerance of the \verb|fsolve| algorithm to iterate until 
Eqs.~\eqref{betheeqns} are solved such that the left hand side equals the right hand side to an absolute accuracy of $10^{-10}$. Once the value of $|c|$ has been increase to be point that the quasi-momenta are within $10^{-10}$ of the string solution the algorithm terminates.

\begin{figure}
\includegraphics[width=7cm]{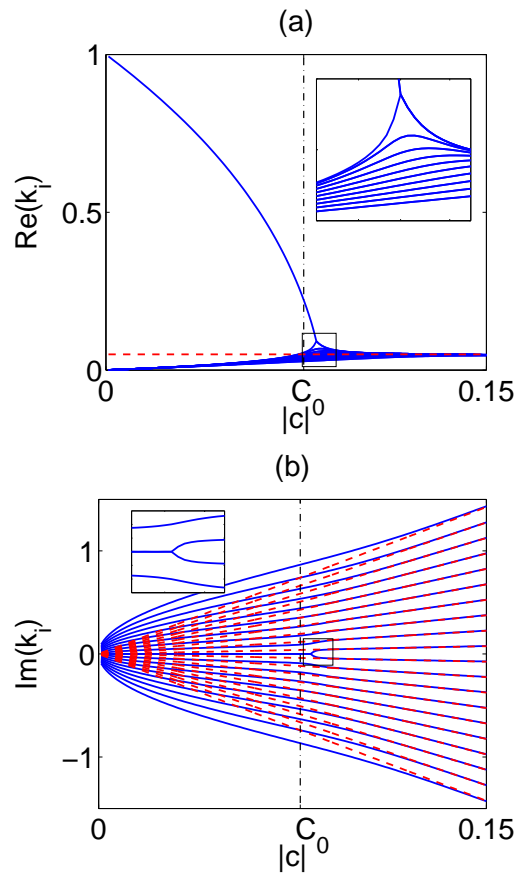}
\caption{{Quasimomenta of the first excited state for $N=20$ particles as as a function of the interaction strengnth $c$. (a) and (b) show the real and imaginary parts respectively of the quasi-momenta. The blue solid lines are the roots calculated using the Bethe ansatz, whereas the red dashed lines show the string solutions given by Eqs.~\eqref{castinguess} which give the asymptotic behaviour of the quasi-momenta in the attractive limit.} The two solutions are in good agreement above the critical interaction strength, validating the essential properties of the \emph{bound states} described in references \cite{castinherzog,mcguire,thacker,laihaus,calogero_degasperis}. The inset shows the region where the quasi-momenta \emph{bifurcate} causing numerical difficulties.}
\label{attractive_limit1}
\end{figure}

\begin{figure}
\includegraphics[width=7cm]{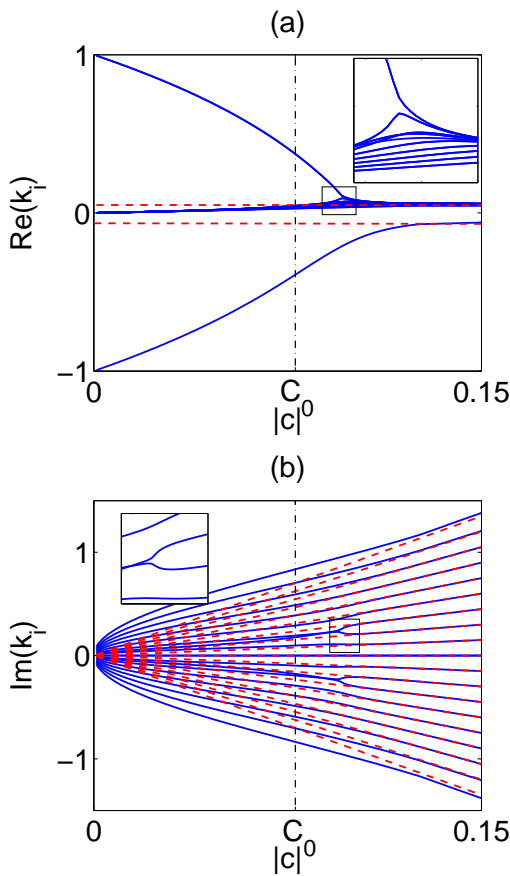}
\caption{{Quasimomenta of one of the second excited states for $N=20$ particles as as a function of the interaction strengnth $c$. (a) and (b) show the real and imaginary parts respectively of the quasi-momenta. The blue solid lines are the roots calculated using the Bethe ansatz, whereas the red dashed lines show the string solutions given by Eqs.~\eqref{attractive_guess1} with $M=1$}
 The inset shows where the quasi-momenta \emph{bifurcate} for this case.}
\label{attractive_limit2}
\end{figure}

\begin{figure*}
\includegraphics[width=17cm]{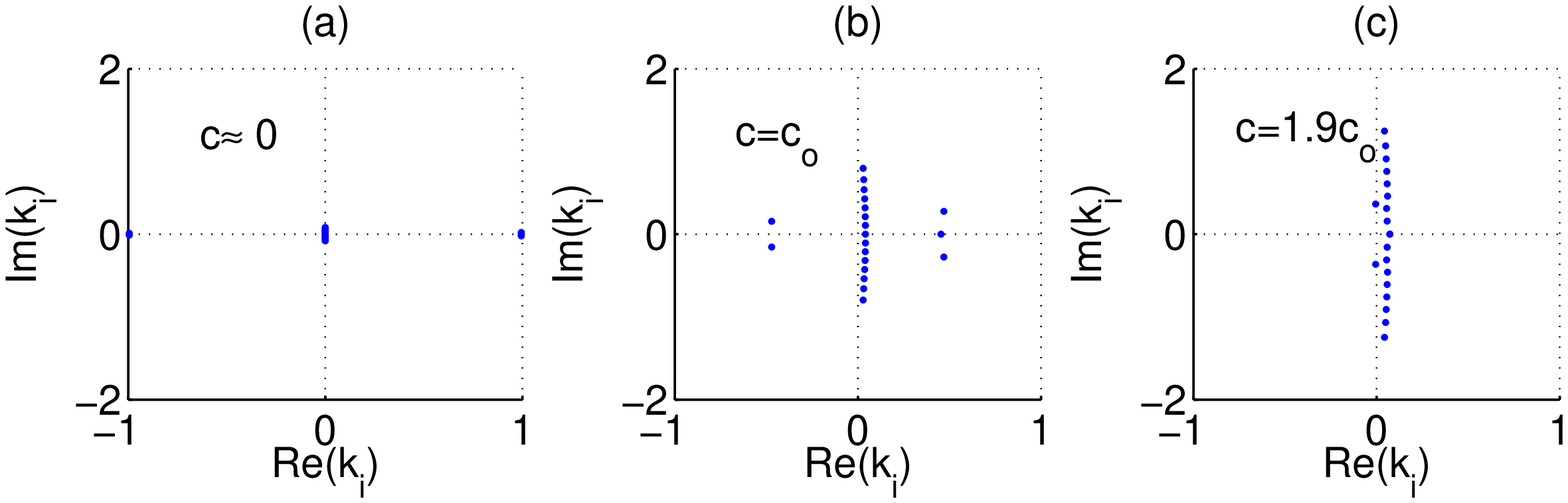}
\caption{{Quasi-momenta distribution for $N=20$ particles in the complex plane for different values of interaction strength $c$ for the first excited state as in Fig.~\protect\ref{attractive_limit1}. (a) $|c|=10^{-6}$. The quasi-momenta are close to the single particle momenta for the ideal gas. (b) $|c|=C_0\approx0.079$.  Near the mean-field critical point there exists a two-particle bound state, a three-particle bound state and a fifteen-particle bound state. (c) $|c|=1.9$. The interaction strength is well past the critical point and the three particle bound state has collapsed in with the fifteen particle bound state, however there is still a two particle bound state.}}
\label{comp1}
\end{figure*}

\begin{figure*}
\includegraphics[width=17cm]{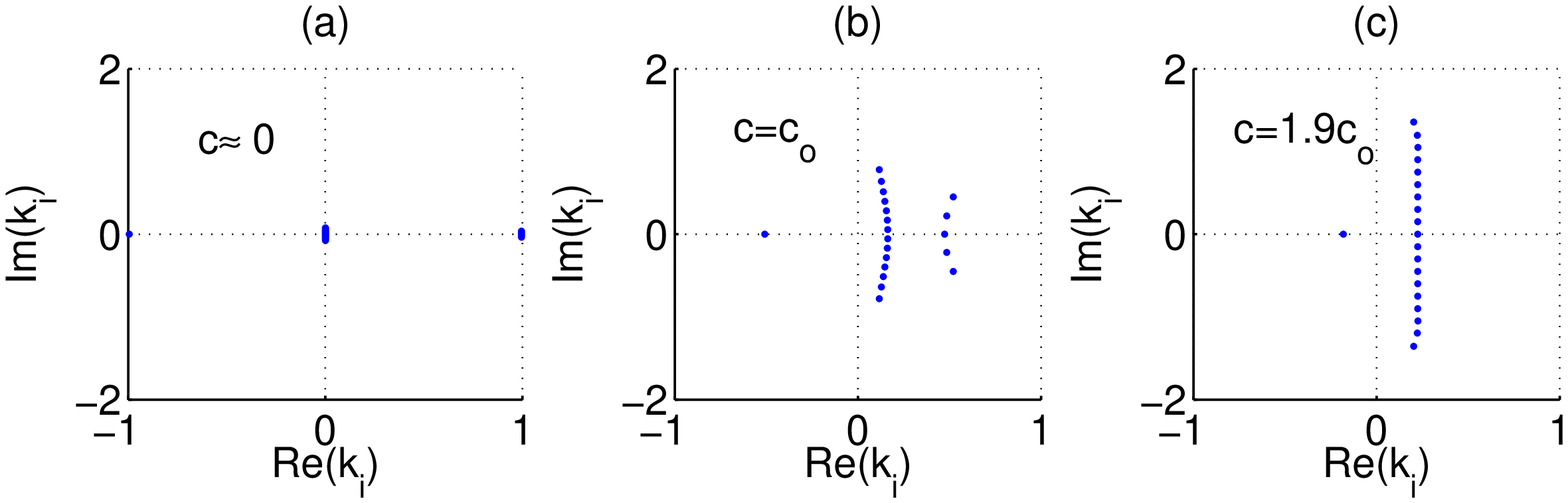}
\caption{{Quasi-momenta distribution for $N=20$ particles in the complex plane for different values of interaction strength $c$ for the first excited state as in Fig.~\protect\ref{attractive_limit2}. (a) $|c|=10^{-6}$. The quasi-momenta are close to the single particle momenta for the ideal gas. (b) $|c|=C_0\approx0.079$.   Near the mean-field critical point, there exists a free particle, a five-particle bound state and a fourteen-particle bound state. (c)$|c|=1.9$. The interaction strength is well past the critical point and the five-particle bound state has combined with the fourteen-particle bound state leaving a single free particle.}}
\label{comp2}
\end{figure*}

We present the results of these calculations in two different ways. Figures \ref{sixpartkatt}, \ref{attractive_limit1}, and \ref{attractive_limit2}  show the continuous evolution of the real and imaginary parts of the quasi-momenta, for a number of excited states as $c$ becomes more negative.  Figure~\ref{sixpartkatt} shows eight excited states for $N=6$, whereas Figs.~\ref{attractive_limit1} and \ref{attractive_limit2} show two representative  excited states for $N=20$. Figures \ref{comp1}, and \ref{comp2} show snapshots of the distribution of the quasimomenta in the complex plane for $N=20$. {Our calculations elucidate  the \emph{deviations} from the string solutions at finite density~\cite{thacker}.}

We see from Fig.~\ref{sixpartkatt}(b) and (e) that as the real parts of $k_i$ and $k_j$ become equal the imaginary parts \emph{bifurcate}, and vice-versa. These splittings of the quasi-momenta provide additional challenges for the root-finding algorithm. Because the splitting often occurs sharply, the initial guess for the quasi-momenta found from the previous value of $c$ can be sufficiently inaccurate that the algorithm does not converge.
 Thus if one finds that the root-finding algorithm is not converging at a particular value of $c$ then this could be an indication that two equal quasi-momenta are beginning to split. When this occurs there is always a degeneracy between the quasi-momenta as to which one \emph{goes up} and which one \emph{goes down} (this degeneracy is of no \emph{physical} consequence, it is merely a mathematical hurdle). The numerical fluctuations of the values of the quasi-momenta is usually somewhere close to machine precision, $\approx10^{-15}$, and are difficult to detect. In order to observe the splitting we found it was necessary to manually alter the initial guess for the quasi-momenta at these points, forcing one to \emph{go up} and one to \emph{go down} (again see Fig.~\ref{sixpartkatt} (b) and (e)).

There is no fundamental limit to the excited states we can reach with this procedure, and similar graphs can be produced for  up to approximately $N=20$ particles (see Fig.~\ref{attractive_limit1} and \ref{attractive_limit2} as examples of the $N=20$ particle gas). 
In principle this work could be extended to larger numbers of particles, and in fact we have found the ground state for up to $N\approx50$. However, the \emph{bifurcations} in the excited states require some time to locate {and we have had to deal with them manually.  This means that} the algorithm begins to take a considerable amount of time for $N$ significantly larger than twenty.

Roots of Eq.~\eqref{betheeqns} for $c<0$ have a far more complicated behaviour than in the more commonly studied repulsive regime. Apart from the obvious reason that the quasi-momenta can be complex (as opposed to strictly real for the repulsive gas~\cite{korepinbook}), there also exists the change in behaviour at the critical interaction strength $C_0$. For some time it has been suspected that the roots for large enough attraction would take values on \emph{strings} in the complex plane which correspond to some bound state of the system~\cite{laihaus,thacker,castinherzog}. That is, the wave function would represent some kind of localised state~\cite{thacker,mcguire} such as
\begin{eqnarray}
\psi_{bound}(\theta_1,\ldots,\theta_N)=\exp\left(-\frac{1}{2}|c|\sum_{\average{i,j}}\lvert \theta_i-\theta_j\rvert\right),\label{boundstate}
\end{eqnarray}
and the corresponding quasi-momenta of this state would be
\begin{eqnarray}
k_1&=&\frac{K}{N}+\frac{1}{2}(N-1)ic,\nonumber\\
k_2&=&\frac{K}{N}+\frac{1}{2}(N-3)ic,\nonumber\\
&\vdots&\nonumber\\
k_N&=&\frac{K}{N}-\frac{1}{2}(N-1)ic,\label{castinguess}
\end{eqnarray}
where $K=\sum k_i$ is the total momentum of the state. It is worthwhile to note that although we refer to the wave function in Eq.~\eqref{boundstate} as a \emph{localised state} it is not localised to any specific point. Rather it has a localised pair correlation~\cite{calabrese_caux1}
\begin{eqnarray} 
g^{(2)}(\theta,\theta')=\frac{\average{\hat{\Psi}^{\dagger}(\theta)\hat{\Psi}^{\dagger}(\theta')\hat{\Psi}(\theta)\hat{\Psi}(\theta')}}{\average{\hat{\Psi}^{\dagger}(\theta)\hat{\Psi}(\theta)}\average{\hat{\Psi}^{\dagger}(\theta')\hat{\Psi}(\theta')}},
\end{eqnarray} 
where $\hat{\Psi}(\theta)$ ($\hat{\Psi}^{\dagger}(\theta)$) are bosonic field operators which annihilate (create) a particle at position $\theta$. Thus measurements of the density of an ensemble will always yield a localised projection, whilst the system is not localised \emph{prior} to the measurement. One can think of the system prior to the measurement as being in a (macroscopic) superposition of localisation at \emph{every} point on the ring. This form for the quasi-momenta given by Eqs.~\eqref{castinguess} is commonly referred to in the literature as a \emph{string} solution~\cite{thacker,castinherzog,laihaus}. It is often assumed that multiple strings may exist for one system (\cite{thacker} and references therein), with each string corresponding to a soliton of different momentum. This choice of quasi-momenta also seems intuitively reasonable and gives the result, mandated by McGuire~\cite{mcguire,thacker} for the total energy of the bound state
\begin{eqnarray}
E_K = \frac{1}{N}K^2-\frac{N(N^2-1)}{12}c^2.\label{mcguireenergy}
\end{eqnarray}
The degree to which the system is localised by these string solutions can be quantified by the absolute values of the imaginary parts of the quasi-momenta. Thus, in Eqs.~\eqref{castinguess}, as $|c|$ increases, so does the degree of localisation (as one would naively expect). However, it is easy to show that in the ideal limit, $c\rightarrow0$, the quasi-momenta are exactly the single particle momentum states. Thus if one accepts the quasi-momenta given by Eqs.~\eqref{castinguess}, then one is left with chronic discontinuities in the behaviour of the quasi-momenta as a function of $|c|$. Our results give direct evidence that the two limits are bridged via Eqs.~\eqref{betheeqns}. We also find other families of excited states, with multiple bound states forming in the limit of strong attraction. These show  agreement with the truncated diagonalization results {discussed later in section \ref{comparison}}. These families of excited states can be interpreted as the formation of multiple bound states. It is very interesting to observe this intermediate behaviour between the ideal gas and strongly attractive gas since this is where soliton formation occurs at zero temperature.

In Fig.~\ref{sixpartkatt} we see the behaviour of the quasi-momenta for a system of $N=6$ particles in the eight lowest energy states ordered via their energies at $c=0$. As $|c|$ becomes larger, energy level crossings are observed as in Fig.~\ref{sixpartbetheatt}. The quasi-momenta begin at $c=0$ at their respective single particle momentum states, and as $c$ becomes more negative the quasi-momenta tend toward a string solution as predicted. The initial behaviour of the imaginary part of the quasi-momenta goes like $\sqrt{\lvert c\rvert}$ for $|c|<C_0$, but for $|c|>C_0$ when the system is in a localised state the low-lying eigenstates have quasi-momenta corresponding to the string solutions Eqs.~\eqref{castinguess}. 

This is verified again for the $N=20$ particle gas in Fig.~\ref{attractive_limit1} where the solid blue lines show the quasi-momenta as a function of $\lvert c\rvert$ and the red dashed lines show the string solution Eqs.~\eqref{castinguess}, and both are in agreement in the strongly attractive limit. However, we also find excited states that do not agree with Eqs.~\eqref{castinguess}, but instead form multiple bound states. This results in a decrease in the degree of localisation. In these cases we find the behaviour of the quasi-momenta at large $|c|$ to be given by 
\begin{eqnarray}
k_1&=&\frac{K}{M}+\frac{1}{2}(M-1)ic\nonumber\\
&\vdots&\nonumber\\
k_{M}&=&\frac{K}{M}-\frac{1}{2}(M-1)ic\nonumber\\
k_{M+1}&=&\frac{K'}{N-M}+\frac{1}{2}(N-M-1)ic\nonumber\\
&\vdots&\nonumber\\
k_N&=&\frac{K'}{N-M}-\frac{1}{2}(N-M-1)ic.\label{attractive_guess1}
\end{eqnarray}
We interpret this solution physically as an $M$ particle bound state with momentum $K$ and an $N-M$ particle bound state with momentum $K'$. 

Evidence of these states can be seen in Fig.~\ref{attractive_limit2} where the solid blue lines show the quasi-momenta as a function of $\lvert c\rvert$, whereas the red dashed lines show the solutions given by Eqs.~\eqref{attractive_guess1} with $M=1$, $K=-1/20$ and $K'=21/380$. 
In Fig.~\ref{sixpartkatt}(d) and (f) the same kind of behaviour is seen for the $N=6$ particle gas. The behaviour is perhaps clearer in Figs.~\ref{comp1}(c) and \ref{comp2}(c) where one can see the two distinct strings in the complex plane. One can derive a simple expression for the energy of these states (similar to Eq.~\eqref{mcguireenergy}),
\begin{eqnarray}
E^{(1)}_{KK'}=&&\frac{(N-M)K^2+M{K'}^2}{M(N-M)}-\nonumber\\
&&\frac{N(N^2-3NM+3M^2-1)}{12}c^2.\label{newenergy}
\end{eqnarray}
This trend continues such that three or more bound states of atoms occur within the system. The evidence for this is most easily seen by the grouping of the quasi-momenta in the complex plane [Figs.~\ref{comp1}(c) and \ref{comp2}(c)].

Thus for this system once $c<C_0$ and localisation has occurred we observe several different families of solutions. The first is the ground state of the system, corresponding to a single, stationary $N$-particle soliton. Then there are the elementary excitations of this state whereby the soliton has some (integer valued) total momentum about the ring. Furthermore there exist the higher order excitations in which multiple solitons form around the ring. Equally these multiple solitons can have elementary excitations of their own, corresponding to some integer valued total momentum.

\subsection{The Excitation Spectrum}\label{exc_spect}

\begin{figure}
\includegraphics[width=8cm]{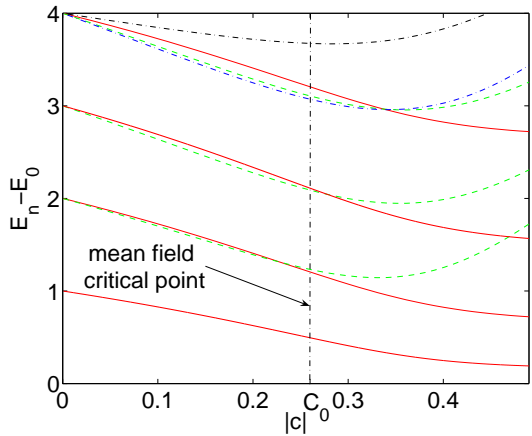}
\caption{(color online) The excitation spectrum of a $N=6$ particle system obtained via the Bethe ansatz. The solid red lines show the formation of a single bound state. The green dashed lines show the formation of two bound states. The blue dot-dash lines show the formation of three separate bound states. The qualitative change in the behaviour of the eigenstates at the phase transition point becomes more visible than for the two particle case (see Fig.~\ref{betheexactcompare}).}
\label{sixpartbetheatt}
\end{figure}

We can now calculate the excited states of the system via the Bethe ansatz, using Eq.~\eqref{energy} and the numerically determined roots of Eqs.~\eqref{betheeqns} (the quasi-momenta). We show a comparison between the truncated diagonalization approach of Kanamoto \emph{et~al.}~\cite{kanamoto1,kanamoto2} in section \ref{comparison} (see Figs. \ref{betheexactcompare} and \ref{betheexactcompare2}). We plot the excitation spectrum for $N=6$ particles in Fig.~\ref{sixpartbetheatt} and $N=20$ particles in Fig.~\ref{energy20}. The point at which mean-field theory predicts the quantum phase transition, $C_0$ (see appendix \ref{meanfieldc0}), is indicated on all figures with a vertical dot-dashed line. We observe the \emph{sharpening} of the cross-over regime as $N$ increases as expected for a phase transition in the limit of large $N$.

\begin{figure}
\includegraphics[width=8cm]{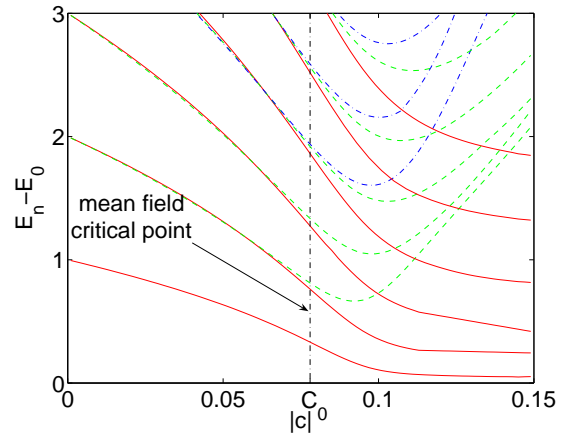}
\caption{ The excitation spectrum of a $N=20$ particle system obtained via the Bethe ansatz. The solid red lines show the formation of a single bound state. The green dashed lines show the formation of two bound states. The blue dot-dash lines show the formation of three separate bound states.}
\label{energy20}
\end{figure}
Once the size of the system gets up to around $N=20$, the difference in the behaviour of the states for $|c|<C_0$ and $|c|>C_0$ is clear. For $|c|>C_0$ transition (once the bound state has formed), the different families of solutions to the Bethe Eqs.~\eqref{betheeqns} (discussed in the previous section) become distinguishable through their separation from the ground state. The solid red lines in Figs. \ref{sixpartbetheatt} and \ref{energy20} show the behaviour of the single bound states found by McGuire and others~\cite{mcguire,castinherzog,laihaus,calogero_degasperis} given by Eqs.~\eqref{castinguess}, the green dashed lines show the family of solutions given by Eqs.~\eqref{attractive_guess1} which are made up of two separate bound states. Finally the blue dot-dashed lines show the existence of three separate bound states within the gas.

\subsection{Comparison to the truncated Hilbert space diagonalization}\label{comparison}
\begin{figure}
\centering
\includegraphics[width=9cm]{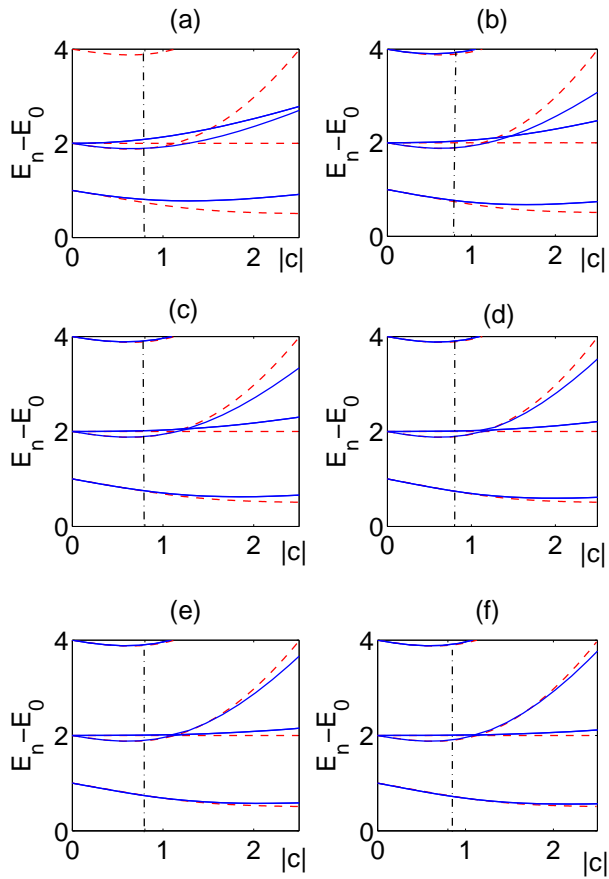}
\caption{Comparison of the excitation spectrum found via the Bethe ansatz Eqs.~\eqref{betheeqns} (red dashed lines) to that found via diagonalization of the truncated Hamiltonian (solid blue lines) for a $N=2$ particle system.
 The vertical dash-dotted line marks the mean field crititical interaction strength $C_0$ derived in appendix \ref{meanfieldc0}.  (a) Single particle momentum state cut-off of $\pm1$. (b) Single particle momentum state cut-off of $\pm2$, and so on, up to (f) which has a single particle momentum cut-off of $\pm6$.}
\label{betheexactcompare}
\end{figure}
\begin{figure}
\includegraphics[width=8cm]{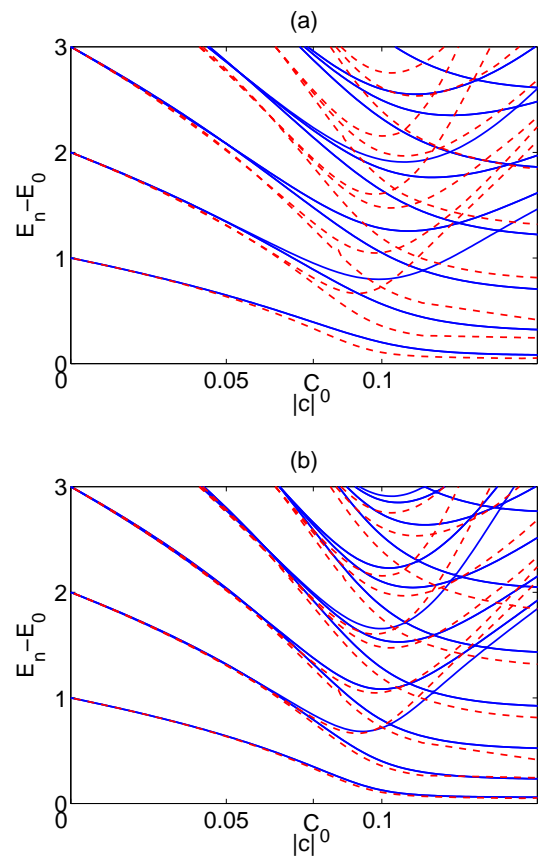}
\caption{Comparison of the excitation spectrum found via the Bethe ansatz Eqs.~\eqref{betheeqns} (red dashed lines) to that found via diagonalization of the truncated Hamiltonian (solid blue lines) for a $N=20$ particle system. (a) Truncation at $m=3$ (single particle momentum states $0,\pm1$). 
(b)Truncation at $m=5$ (single particle momentum states $0,\pm1,\pm2$). }
\label{betheexactcompare2}
\end{figure}

The excitation spectrum of the attractive gas has been numerically studied via truncation of the Hilbert space in previous work~\cite{kanamoto1,kanamoto2}. The key element in the approach is to exclude all single particle momentum states beyond a particular threshold. The results of these calculations are of course exact at zero interactions, but deviate from the exact solution as the interaction strength, $|c|$, is increased. The degree of deviation can of course be decreased by increasing the threshold momentum state, however the dimension of the Hilbert space will eventually become unmanageably large in doing so. An example of this process is shown in Fig.~\ref{betheexactcompare} for a two particle system where it is possible to make the cut-off relatively high without having the Hilbert space dimension expand beyond a practical number.
The results indicate a good qualitative description from a very small number of momentum states~\cite{kanamoto1}. For example if $N=200$, a momentum cutoff of $\pm1$ is reasonable up to and even beyond ($c\approx2c_0$) the critical point. This was supported by calculations made by Kavoulakis~\cite{kavoulakis} who used a mean field approximation with a suitably chosen wave function to examine the ground state and low-lying states near the phase transition.

The truncated diagonalization procedure we use follows that of Kanamoto \emph{et~al.}~\cite{kanamoto2}. The first quantised Hamiltonian in Eq.~\eqref{firstham} is rewritten in second quantised formalism as,
\begin{eqnarray}
\hat{H}=\int_0^{2\pi}d\theta\left[-\hat{\Psi}^{\dagger}(\theta)\pderivtwo{}{\theta}\hat{\Psi}(\theta)+c\hat{\Psi}^{\dagger2}(\theta)\hat{\Psi}^2(\theta)\right],\label{secondham}
\end{eqnarray}
where $\hat{\Psi}(\theta)$ is the bosonic field operator that annihilates a boson at coordinate $\theta$ and obeys the usual commutation rules and the periodic boundary conditions of the ring geometry. This field operator is then approximated by its truncated expansion in terms of single particle states
\begin{eqnarray}
\hat{\Psi}(\theta)=\sum_{j=-k_0}^{k_0}\varphi_j(\theta)\hat{c}_j,\label{expansion}
\end{eqnarray}
where $\varphi_j(\theta)$ is the single particle eigenstate with angular momentum $j$ (see appendix \ref{singleparticle}), $\hat{c}_j$ annihilates a boson with angular momentum $j$ and $k_0$ is the single particle momentum state cut-off. In this manner, for finite $N$ and finite $k_0$, $\hat{H}$ is a finite matrix. Furthermore, $\hat{H}$ (when written in the appropriately ordered basis set) will be block diagonal, as it can be divided into total momentum subspaces which can be individually diagonalised. If we use $m$ to denote the total number of single particle momentum states used in the expansion (i.e. $m=2k_0+1$) then the total dimension of the Hilbert space will be the binomial coefficient $\binom{N+m-1}{N}$. This Hilbert space will split up into $N(m-1)+1$ (total momentum) subspaces. The largest subspace will be the zero momentum subspace.
 For example if we were to consider a gas of $N=20$ particles with the momentum cut-off at $\pm2$ ($m=5$), then the total dimension of the truncated Hilbert space would be 10626. However this splits up into 81 different total momentum subspaces, the largest of which has dimension 318, which can be exactly diagonalised on a standard desktop PC. Table \ref{hilbertdiminfo} gives an indication of the scale of computation required for the truncated diagonalization procedure.

In Fig.~\ref{betheexactcompare2} we show a comparison between excitation spectrums obtained from the truncated diagonalization procedure and the Bethe ansatz for the $N=20$ particle gas. We observe that for $m=3$ (single particle momentum cut-off at $\pm1$) the truncated diagonalization over-estimates the energy difference between ground and excited states in the cross-over regime, but under-estimates the difference in the regime where the bound states have formed. For $m=5$ (single particle momentum cut-off at $\pm2$) the agreement between the two methods in the cross-over regime is vastly improved, however the truncated diagonalization still underestimates the difference between ground and excited states in the bound state regime.

\begingroup
\squeezetable
\begin{table}
\begin{ruledtabular}
\begin{tabular}{ccccc}
\bf{N}&\bf{m}&\bf{Hilbert space}&\bf{Number of}&\bf{Dimension of} \\
 & &\bf{dimension}&\bf{subspaces}&\bf{largest subspace}\\
\hline
 2 & 3 & 6  & 5  & 2 \\
   & 5 & 15 & 9  & 3 \\
   & 7 & 28 & 13 & 4 \\
\hline
20 & 3 & 231 & 41 & 11 \\
   & 5 & 10626 & 81 & 318 \\
   & 7 & 230230 & 121 & 5444 \\
\hline
200 & 3 & 20301 & 401 & 101 \\
    & 5 & 70058751 & 801 & 230673 \\
\end{tabular}
\end{ruledtabular}

\caption{The size of the Hilbert space and subspaces for selected particle numbers and momentum cut-offs. This gives an idea of the size of computation required for the truncated diagonalisation procedure.}
\label{hilbertdiminfo}
\end{table}

\endgroup

\begin{figure*}
\includegraphics[width=18cm]{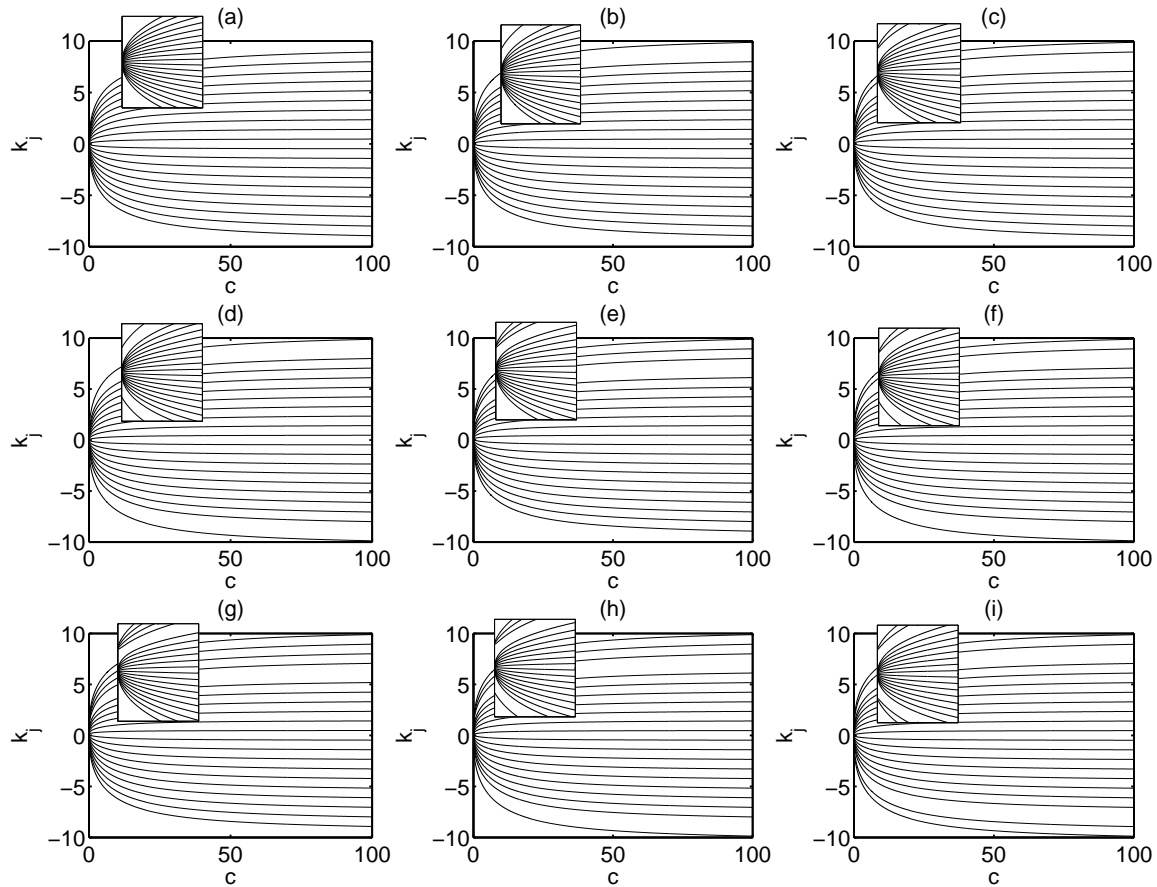}
\caption{The quasi-momenta as a function of interaction strength for $c>0$. (a) Ground state of the ideal Bose gas. (b) First excited state of the ideal Bose gas (degenerate with the state $k_j\rightarrow-k_j$). (c), (d) Degenerate second excited states of the ideal Bose gas (also degenerate with the state $k_j\rightarrow-k_j$). (e), (f) degenerate third excited states of the ideal Bose gas (also degenerate with the state $k_j\rightarrow-k_j$). (g), (h) , (i) Degenerate fourth excited states of the ideal Bose gas (also degenerate with the state $k_j\rightarrow-k_j$). The insets on (a)-(i) show the behaviour close to $c=0$.}
\label{repulsivek}
\end{figure*}

\section{The Repulsive Gas: $c>0$}
\label{sec:repulsive}

The root-finding algorithm can easily be applied to the case of repulsive attractions. A proof exists showing that all roots of Eqs.~\eqref{betheeqns} lie on the real axis \cite{korepinbook}. This means that we avoid the problems associated with the bifurcations which appeared in the attractive case and the situation is therefore significantly easier.
In this case the limiting behaviour as $c\rightarrow\infty$ of the quasi-momenta is given by the single particle momentum states of an ideal Fermi gas~\cite{girardeau,liebliniger}. This limit is known as the Tonks-Girardeau gas and has been experimentally observed~\cite{tonksgas}.

In this work we use our numerical root-finding algorithm to bridge from the ideal gas to the Tonks-Girardeau gas. We do so for both the ground state and the low-lying excited states for finite numbers of particles. The ground state has appeared previously in the work of Sakmann \emph{et al.}~\cite{sakmann}. The thermodynamic limit has been addressed in the homogeneous case~\cite{lieb,yangyang1,yang1} and the trapped case~\cite{petrov1} (and references therein).

\subsection{The Quasi-Momenta}

The root-finding algorithm is essentially unchanged from the case of attractive interactions. It is worth mentioning however that, since there is no phase transition in the repulsive gas the step size $\Delta c$ can be increased without sacrificing a reasonable initial guess. In our calculations we use $\Delta c=10^{-4}$ for $0\leq c\leq1$, then $\Delta c=10^{-3}$ for $1\leq c\leq6$, then $\Delta c=10^{-2}$ for $6\leq c\leq20$, then $\Delta c=10^{-1}$ for $20\leq c\leq 100$ and finally $\Delta c=1$ for $c\geq100$.

In Fig.~\ref{repulsivek} we see how the quasi-momenta behave as a function of the interaction strength $c$. The single particle excitations of the ideal \emph{bosons} at $c=0$ result in single particle excitations of the ideal \emph{fermions} at $c=\infty$. The concept of particle-hole excitations for the strongly repulsive gas (see references~\cite{yangyang1,thacker}) is clear in Fig.~\ref{repulsivek}. We see low-lying excited state quasi-momenta are similar to that for the ground state, but with a small number of quasi-momenta excited, leaving behind \emph{holes}.

\begin{figure}
\includegraphics[width=8.6cm]{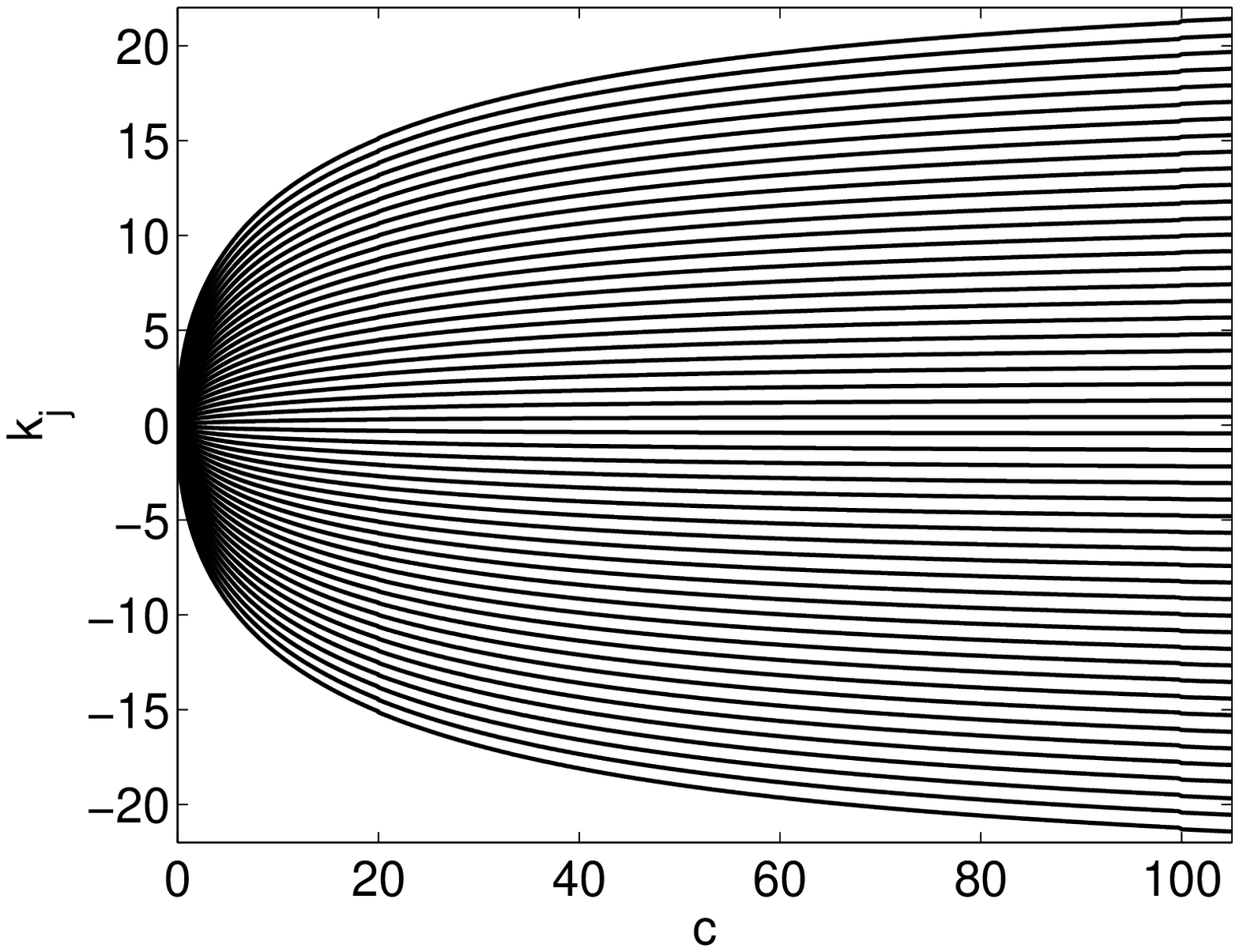}
\caption{The quasi-momenta as a function of interaction strength for an $N=50$ particle repulsive gas.}
\label{largeN}
\end{figure}

It is possible to push the algorithm for the repulsive case to much higher atom numbers than with the attractive case. This is due to the absence of the exotic features in the quasi-momenta spectrum (see for example the splitting occurring in the insets of Figs \ref{attractive_limit1} and \ref{attractive_limit2}). In Fig.~\ref{largeN} we show the ground state quasi-momenta for a repulsive gas with $N=50$.
\subsection{The Excitation Spectrum}

\begin{figure}
\includegraphics[width=8.6cm]{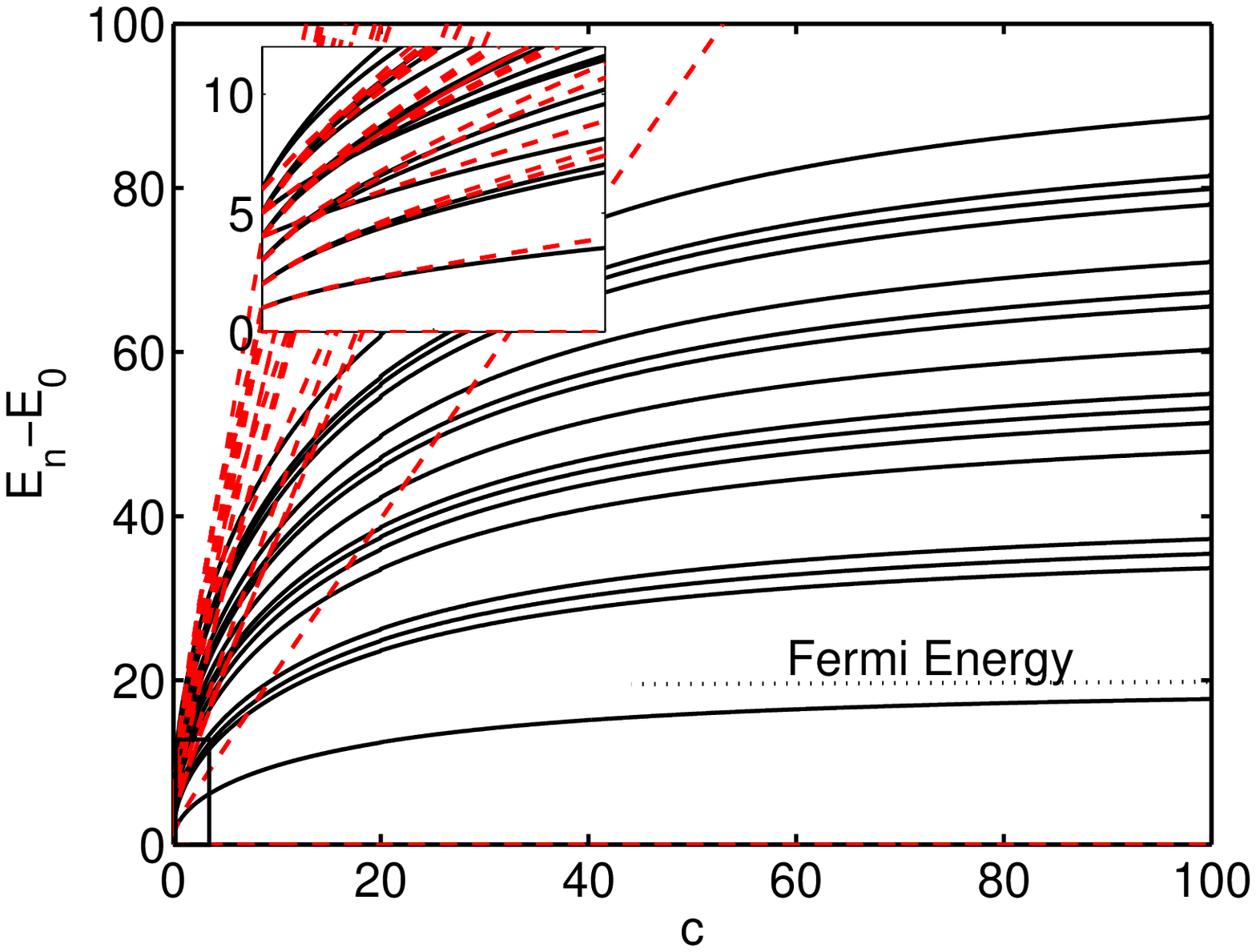}
\caption{The  excitation spectrum of a $N=20$ particle gas with repulsive interactions $c>0$ (solid black lines.  The red dashed lines were obtained by truncating the Hilbert space down to five single particle momentum states. The inset shows the region of $0<c<1$ where the truncated diagonalization is accurate. The dotted lines shows the ground state energy of a free Fermi gas, and the so-called Tonks Girardeau ground state asymptotes to this value.}
\label{repulsiveenergy}
\end{figure}

As for the attractive gas, the excitation spectrum can be found using Eq.~\eqref{energy}. In  Fig.~\ref{repulsiveenergy} we show the energies of the low lying excited states for $c>0$. The red dashed lines show the result obtained from truncating the Hilbert space down to five single particle states and diagonalising the Hamiltonian (see section \ref{comparison}). The truncation procedure is accurate for small enough values of $c$ (see the inset of Fig.~\ref{repulsiveenergy} where $c$ runs from zero to one), but fails as $c$ increases beyond one. The dotted line shows the ground state energy of a $N=20$ particle Fermi gas.

\section{Conclusions}
\label{sec:conclusions}
We have studied the behaviour of the roots of the Bethe ansatz Eqs.~\eqref{betheeqns} for the 1D Bose gas on a ring as a function of the interaction strength for both the attractive case (up to $N=20$ particles) and the repulsive case. We used these roots to determine the exact many-body energy spectrum of the one-dimensional Bose gas with $\delta$-function interactions in a 1D ring trap for small numbers of particles. We compared our results to those obtained via the approximate method of truncated diagonalization, to which we found good qualitative agreement and reasonable quantitative agreement for small enough interaction strength.  
We have quantitatively addressed the issue of when it is reasonable to assume a bound state solution to Eqs.~\eqref{betheeqns}, and have quantified the deviations from the string solutions that occur for a finite density gas, thus establishing the point at which the string solutions become valid.
Furthermore we found evidence for the existence of
multiple, independent \emph{solitons} existing on the ring. These excitations had been hypothesised in previous work, however to the best of our knowledge this is the first direct evidence of their existence. 
We have described the analytical expressions for the quasi-momenta of these new families of string solutions  in Eqs.~\eqref{attractive_guess1}) and an expression for the energies of these states.  
It may be interesting in the future to make use of the solutions found in this paper to calculate correlation functions for the one-dimensional Bose gas.

\section{Acknowledgements}
This work was supported by the Australian Research Council Centre of Excellence for Quantum-Atom Optics and the Queensland State government. The authors wish to acknowledge the useful discussions with all members of the University of Queensland node. AGS wishes to particuarly thank  Chris Foster and David Barry for technical help with computational tasks and J.~S.~Caux and Norman Oelkers for useful discussions.

\begin{appendix}

\section{Single particle in a ring}\label{singleparticle}
The single particle Hamiltonian is
\begin{eqnarray}
-\frac{\hbar^2}{2mR^2}\pderivtwo{}{\theta}\varphi(\theta)=E\varphi(\theta)\label{singleham},
\end{eqnarray}
and the single particle energy eigenvalues are
\begin{eqnarray}
E_k=\frac{k^2\hbar^2}{2mR^2},\qquad k=0,\pm1,\pm2\ldots,\label{singleenergy}
\end{eqnarray}
with corresponding eigenstates are
\begin{eqnarray}
\varphi_k(\theta)=\frac{1}{\sqrt{2\pi}}e^{ik\theta}.\label{singlestate}
\end{eqnarray}

\section{The mean-field critical point}\label{meanfieldc0}
We include a derivation of the mean-field critical point of localisation using the same approach as Kavoulakis~\cite{kavoulakis}. Considering the Hamiltonian \eqref{secondham} we define a mean field $\psi(\theta)$ and approximate the system as having the many body wave function $\prod_i\psi(\theta_i)$. We next expand $\psi(\theta)$ in terms of single particle states given by Eqs.~\eqref{singlestate}, however we truncate all momentum states greater than one,
\begin{eqnarray}
\psi(\theta)=\alpha_{-1}\varphi_{-1}(\theta)+\alpha_0\varphi_0(\theta)+\alpha_1\varphi_1(\theta).\label{orderparam}
\end{eqnarray}
We then assume $|\alpha_0|\gg|\alpha_{-1}|,|\alpha_1|$ and from the symmetry $|\alpha_{-1}|=|\alpha_{1}|$, and finally the normalisation condition $|\alpha_{-1}|^2+|\alpha_0|^2+|\alpha_1|^2=1$. The energy per particle $\epsilon$ is then found to be
\begin{eqnarray}
\epsilon=&&2|\alpha_1|^2+\frac{\gamma}{2}\left[|\alpha_0|^4+|\alpha_{-1}|^4+|\alpha_1^4|+4|\alpha_0|^2|\alpha_{-1}|^2\right.\nonumber\\
&&+4|\alpha_0|^2|\alpha_1|^2+4|\alpha_{-1}|^2|\alpha_1|^2+2\alpha_0^2\alpha_{-1}^*\alpha_1^*\nonumber\\
&&\left.+2(\alpha_0^2)^*\alpha_{-1}\alpha_1\right],\label{energyperparticle}
\end{eqnarray}
where $\gamma=Nc/\pi$. Setting $\alpha_j=|\alpha_j|e^{i\phi_j}$, then $\epsilon$ is minimised by choosing $\phi_{-1}+\phi_1-2\phi_0=0$, the normalisation condition further simplifies Eq.~\eqref{energyperparticle} to
\begin{eqnarray}
\epsilon-\frac{\gamma}{2}=2|\alpha_1|^2(1+2\gamma)-7\gamma|\alpha_1|^4.\label{energyperparticle2}
\end{eqnarray}
From Eq.~\eqref{energyperparticle2} it is straightforward to see that the approach one must take in order to minimise $\epsilon$ will depend on whether $-1/2<\gamma<0$ (in which case the gas is delocalised) or $\gamma<-1/2$ (in which case the gas is localised). Thus there exists a critical interaction strength of 
\begin{eqnarray}
C_0=\frac{\pi}{2N}.\label{meanfieldpoint}
\end{eqnarray}

\end{appendix}


\end{document}